\documentclass[iop,apj,tighten]{emulateapj}
\usepackage[utf8x]{inputenc}
\usepackage{graphicx}
\usepackage{textcomp}
\usepackage{bm}
\usepackage{amsmath}

\slugcomment{To appear in the Astrophysical Journal.}
\shorttitle{Kinetic helicity and magnetic field in the solar photosphere}
\shortauthors{Sangeetha and Rajaguru}
\begin{document}
\title{Relationships between fluid vorticity, kinetic helicity and magnetic field at the small-scale (quiet-network) 
        on the Sun}
\author{C.R. Sangeetha and S.P. Rajaguru}
\affil{Indian Institute of Astrophysics, Koramangala II Block, Bangalore 560034, India}
\email{rajaguru@iiap.res.in,crsangeetha@iiap.res.in}

\begin{abstract}
We derive horizontal fluid motions on the solar surface over large areas covering the quiet-Sun magnetic network
from local correlation tracking of convective granules imaged in continuum intensity and Doppler velocity 
by the Helioseismic and Magnetic Imager (HMI) onboard the Solar Dynamics Observatory (SDO). From these we calculate 
horizontal divergence, vertical component of vorticity, and kinetic helicity of fluid motions. We study the 
correlations between fluid divergence and vorticity, and that between vorticity (kinetic helicity) and magnetic 
field. We find that the vorticity (kinetic helicity) around small-scale fields exhibits a hemispherical pattern 
(in sign) similar to that followed by the magnetic helicity of large-scale active regions (containing sunspots).
We identify this pattern to be a result of the Coriolis force acting on supergranular-scale flows (both the 
outflows and inflows), and is consistent with earlier studies using local helioseismology. Further, we show 
that the magnetic fields cause transfer of vorticity from supergranular inflow regions to outflow 
regions, and that they tend to suppress the vortical motions around them when magnetic flux densities exceed about 
300 G (HMI). We also show that such action of magnetic fields leads to marked changes in the correlations between 
fluid divergence and vorticity. These results are speculated to be of importance to local dynamo action if present, 
and to the dynamical evolution of magnetic helicity at the small-scale.
\end{abstract}


\section{Introduction}

Interactions between turbulent convection and magnetic field at the photospheric layers play central roles 
in structuring and driving varied forms of dynamical phenomena in the atmospheric layers above, and hence in the 
energetics (see, {\it e.g.} \citet{Nordlund2009} and references therein). These interactions in the near-surface 
layers are also basic to local dynamo action \citep{Schussler2008}, which, if present, could explain the large 
amount of quiet-Sun magnetic flux inferred from high-resolution observations \citep{Lites2008,Goode2010}. 
An important aspect of these interactions is the role of helical or swirly fluid motions which 
can similarly twist or inject helicity to the magnetic field, and vice versa. Helicity of a vector field, in general, 
is defined as the volume integral of the scalar product of the field vector and its curl (or rotation) and it 
quantifies the amount of twistedness in the vector field \citep{Berger1984}. For fluid flow, the kinetic helicity, 
$H_{k}$, is such a quantity derived from velocity $\mathbf{v}$ and its curl or vorticity, 
$\mathbf{\omega} = \mathbf{\bigtriangledown} \times \mathbf{v}$: $H_{k} = \int \mathbf{v}\cdot \mathbf{\omega} dV$, 
where $dV$ is the volume element. For magnetic fields, helicity can be 
calculated in two different ways \citep{Pevtsov1995}: the magnetic helicity, in general, is obtained by applying 
the above definition on the vector potential $\mathbf{A}$ and its curl ({\it i.e.}, the magnetic field 
$\mathbf{B} = \mathbf{\bigtriangledown} \times \mathbf{A}$), $H_{m} = \int \mathbf{A}\cdot \mathbf{\bigtriangledown} \times \mathbf{A} dV$; 
and use of magnetic field and its curl in the above definition gives the so called current helicity, 
$H_{c} = \int \mathbf{B}\cdot \mathbf{\bigtriangledown} \times \mathbf{B} dV$. Both $H_{m}$ and $H_{c}$ 
are measures of twistedness in magnetic field, and they normally preserve signs over a volume of physical interest 
\citep{Seehafer90,Pevtsov1995}. 
Interactions between kinetic and magnetic helicities play fundamental roles in magnetic field generation 
(or dynamo action) as well as in the magneto-hydrodynamical evolution of the fluid and magnetic field \citep{Parker55, Moffat78,
KrauseRadler80, Brandenburg2005}. 

Helicities (magnetic, and kinetic) of solar active regions have been extensively studied using observations of magnetic and 
velocity fields in and around them: a well known property of the active region magnetic fields is the hemispheric 
sign rule, originally discovered by \citet{Hale1927} (see also, \citet{Seehafer90,Pevtsov1995}): active regions in the northern 
hemisphere show a preferential negative magnetic helicity while those in the southern hemisphere show positive 
helicity. The origin or exact cause of such pattern in large-scale magnetic helicity is still not fully understood.
Dynamo mechanisms that impart helicity while the field is being generated as well as transfer of kinetic helicity 
from fluid motions to the magnetic field as it rises through the convection zone \citep{Longcope1998} or 
during and after its emergence at the surface (photosphere) are thought to play roles in the observed pattern \citep{Liu2014b}. 
On the small scale {\footnote{``Large scale'' and ``small scale'' are defined, for the purpose of this paper, to represent the
spatial sizes of coherent magnetic stuctures: large magnetic strcutures such as sunspots are ``large
scale'', while the smaller structures out-lining supergranular boundaries are ``small scale". It is to be noted, however, that these
small magnetic structures are distributed on a large scale all over the solar surface.}}, away from the active and emerging 
flux regions, the magnetic and kinetic helicities
and their interactions are even more poorly understood as measuring them reliably poses significant difficulties. 
Although significant advances have been made in mapping horizontal velocities through correlation tracking 
of surface features such as granulation or magnetic structures ({\it e.g.}, see \citet{Welsch2007} and
references therein), significant uncertainties remain in measuring horizontal components of vector magnetic field 
at the small-scale \citep{Hoeksema2014,Liu2014a} and hence in estimating reliably the current or magnetic
helicities there [Y. Liu and A. Norton -- private communication; {\it cf.} further discussion below.]. 
Observational studies by \citet{Duvall2000,Gizon2003} have explored 
vertical vorticities associated with supergranular scale flows and such results have guided some theoretical 
studies of the relations between kinetic and magnetic helicities in the context of turbulent dynamo 
mechanisms \citep{Rudiger2001,Rudigeretal2001}.

Apart from the above described aspects of interactions between fluid motions and magnetic field, recently, 
vortex motions around small-scale magnetic flux tubes and transfer of helicity from fluid motions to 
magnetic fields have been identified as key players in the upward energy transport and thus in the heating of 
solar corona \citep{Bohm2012}. While \citet{Bohm2012} find vortex flows of life-time of about an hour, 
numerical simulations of \citet{Shelyag2013} show no long-lived vortex flows in the solar 
photosphere. Detection of vortex flows at the granular scale in the photosphere  
date back to the studies by \citet{Brandt1988} and \citet{Simon1989}, who infered that such motions could be 
common features in the granular and supergranular inflow regions. A slightly excess correlation between 
negative divergence of the horizontal flows (or inflows) and vertical vorticity was found by \citet{Wang1995}. 
\citet{Bonet2008} detected a lot of small vortices in the inflow regions and found a clear association 
between them and magnetic bright points. \citet{Innes2009} have detected vorticites in the 
inflow regions by calculating horizonatal flows using balltracking technique. \citet{Balmaceda2010} 
detected strong magnetic flux at the centers of the vortex flows. Vortical flow maps in the quiet-Sun 
were calculated by \citet{Komm2007} and circular flow component of the inflows around active regions were 
calculated by \citet{Hindman2009} using helioseismic ring-diagram technique. A recent work has compared spatially 
resolved veritical vorticites calculated from two independent techniques, local correlation tracking and 
time-distance helioseismology \citep{Langfellner2015}.
 
Despite a good number of studies on vortex flows themselves, there have not been detailed analyses of 
relationships between such fluid motions and magnetic field in the small-scale. For example, there have 
not been reliable inferences on the connections between helicities of fluid motion and magnetic field, and on 
the back-reaction of magnetic field on the fluid.
Much of the difficulties lie in reliably measuring the $H_{m}$ or $H_{c}$ of the small-scale magnetic fields 
as vector field measurements are often subject to large uncertainties outside of sunspots or active regions \citep{Hoeksema2014,Liu2014a}. 
For these reasons, there are no reliable measurements to ascertain if the helicity of small-scale magnetic 
fields follow the hemispheric sign pattern obeyed by active regions. There are conflicting findings regarding 
the dominant signs of current helicity in the weak or small-scale fields over the hemispheres 
\citep{Zhang2006,Sanjay2013}.
Helioseismology results on supergranular-scale flows, however, show the 
effect of Coriolis force introducing a hemispheric sign pattern in the vorticity (and hence kinetic helicity) of 
such flows \citep{Duvall2000,Gizon2003,Gizon2005,Komm2014,Langfellner2015}.

In this work, we focus on examining how the magnetic field modifies the relationships among the fluid dynamical 
quantities, divergence, vorticity and kinetic helicity on the one hand, and how these quantities themselves 
scale against the strength of the magnetic field on the other. Such an analysis is facilitated by the continuous  
full-disk coverage of the Sun in velocity (Doppler), granulation (continuum intensity), and magnetic field 
provided by the Helioseismic and Magnetic Imager (HMI) onboard the Solar Dynamics Observatory (SDO; \citet{Schou2012}). 
Though the spatial resolution of about 1 arc-sec (about 720 km) provided by HMI enables us to track the granulation 
features in both the continuum and velocity images and to derive the horizontal flow field, it is not sufficient 
to resolve the sub-granular scale vortex flows that possibly surround the tiny magnetic flux concentrations. 
Hence, the vertical vorticity that we measure from HMI data through local correlation tracking (LCT) of 
granular motions would have contributions mainly from vortical flows of the size of several granules. Since such 
flows are likely of preponderance at the supergranular boundaries and junctions, we, in our analyses here,
are able to study also the effects of Coriolis force \citep{Duvall2000,Gizon2003} and their influence on the relations between 
vortex motions and magnetic fields.
We discuss the data and the analysis methods in Section 2, results in Section 3, and discussions, conclusions 
and future studies in Section 4.

\section{Data and analysis method}

\begin{figure*}[ht]
\centering
\epsscale{1.05}
\plotone{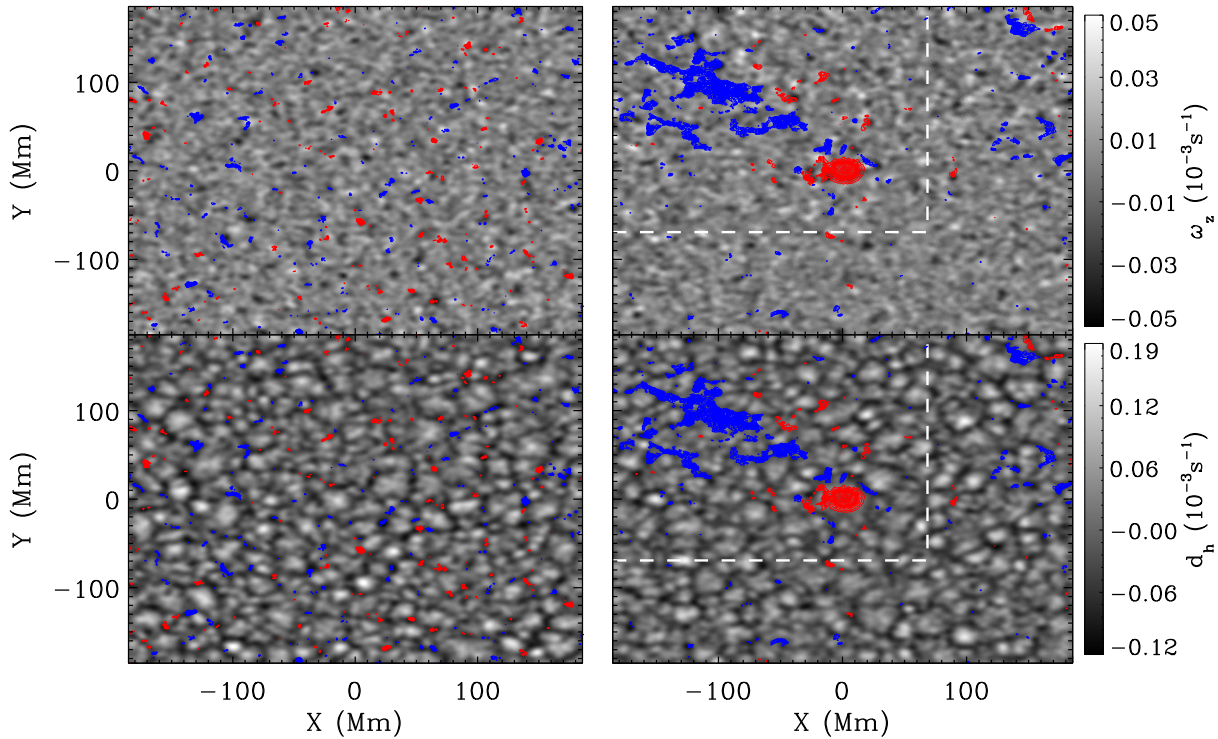}
\caption{\footnotesize Spatial maps of 14 hr averaged vertical vorticities (top panels) and horizontal divergences (bottom panels)
derived from LCT of HMI Doppler velocities.
The left panels show a region consisting of mixed-polarity quiet-network magnetic fields observed on November 3, 2010 with
map center at latitude 15\textdegree S and longitude 0\textdegree; the right panels show a sunspot region observed on August 3, 2010
centered at latitude 15\textdegree N and longitude 0\textdegree.
Contours of magnetic field averaged the same way are overplotted showing field values above $\pm$ 10 G. The red and blue contours correspond
to negative and positive magnetic polarity magnetic fields, respectively. The white dotted lines on the right panels 
separate the sunspot and plage areas from the quiet-network, which is used in the work presented here. }
\label{contourplot}
\end{figure*}

We have used the three major observables from the Helioseismic and Magnetic Imager (HMI) onboard 
the Solar Dynamics Observatory (SDO): Doppler velocities ($v_{d}$), continuum intensities ($I_{\rm{c}}$), 
and line-of-sight magnetic fields $B_{\rm{LOS}}$ (hereafter, we denote $B_{\rm{LOS}}$ simply as $B$). 
The basic datasets are cubes of above observables over about nineteen large regions, typically of size 
30.7 $\times$ 30.7$^{\circ}$ (in heliographic degrees, or $373\times373 \ Mm^{2}$ with 0.03$^{\circ}$/pixel)
covering both northern and southern hemispheres in the latitude range $\pm$30\textdegree and
about $\pm$15\textdegree in longitude about the central meridian, 
tracked for 14 hours and remapped (Postel projected) to a uniform pixel 
size of 0.5$\arcsec$ per pixel. The total area covered by the nineteen regions on the solar surface is 
$19\times 373\times373 \ Mm^{2}$ = $2.65\times10^{6} Mm^{2}$, which is about 0.87 solar hemispheres. 
The nineteen regions chosen are from eleven dates distributed over a period of
years 2010 - 2012. On each of the eight dates, July 11, August 3, October 8, and November 3 of 2010, and
February 8, February 19, July 3, and October 2 of 2011, we have two regions, one centered at 15\textdegree N and
the other centered at 15\textdegree S, and with central longitudes within about 15\textdegree~ of the central meridian. 
On the dates May 8, 2011, June 21, and July 2, 2012, three regions centered at 0\textdegree~  latitude and central 
meridian were included.
Thus the included regions cover equal amount of northern and southern hemispherical areas.
Of the nineteen regions, fourteen are quiet-network regions chosen by examining the magnetograms for the presence of
mixed-polarity network field well away from active regions. The remaining five regions on the five dates,
August 3, 2010, February 19, July 3, October 2, 2011, and July 2, 2012, however, have a sunspot in each; data of the first four regions were available with us already 
and had been used in a different analysis published by the second author \citep{Rajaguruetal2013}. We included
quiet areas of these regions by carefully excluding the sunspots and surrounding plages (one example
is shown Figure 1) covering about 8\textdiscount~of the total area, and hence the total quiet-network area included in
the analysis is about 0.8 solar hemispheres. Since this discarded area of about 0.07 solar hemispheres is all from the
north, we have about 15\textdiscount~excess southern hemispherical area over that of north in the analyses here.
Figure \ref{contourplot} shows two examples from among the analysed regions: 
derived flow divergence and vorticity maps overlaid with magnetic field contours (refer to the 
following section) over a mixed-polarity quiet-network area (left panels) and over a region covering a sunspot and plages (right panels).
The white-line boundaries in the right panel of Figure \ref{contourplot} separate the quiet-Sun areas included in the analysis 
for this region, and these were chosen by visually examining the magnetograms to avoid sunspot and surrounding plages and to 
include only the quiet-Sun network. These straight-line boundaries are just for convenient and easy inclusion of the chosen areas 
in the analyses.

We apply the Local Correlation Tracking (LCT) technique \citep{November1988} on $v_{d}$ and $I_{\rm{c}}$ to 
derive horizontal motions of convective granules. We use the code FLCT \citep{Welsch2004} that implements 
LCT through measurement of correlation shifts in the Fourier space. FLCT is applied on two images separated 
by a time $\bigtriangleup t$. Each image is multiplied by a Gaussian of width $\sigma$ centered at the pixel where 
velocity has to be derived. Cross-correlation is done within this Gaussian window
to calculate the shifts in x- and y-directions that maximise the correlation. The shifts in x- and y-directions are 
divided by $\bigtriangleup t$ to get velocities in x- and y-directions. We remove the $f$ (surface gravity) and
$p$ mode oscillation signals in $v_{d}$ and $I_{\rm{c}}$ before applying the LCT to derive fluid motions. This is done
using a Gaussian tapered Fourier frequency filter that removes frequency components above $1.2~mHz$.
The FWHM of Gaussian window for LCT is $\sigma=15$ pixels and  
$\bigtriangleup t$ is about 2 minutes. We apply FLCT at every time-step, {\it i.e.} every 45 seconds, to derive the 
horizontal velocity components $v_{x}(x,y,t)$ and $v_{y}(x,y,t)$ with the original resolution as the data. From 
these horizontal components of velocity, we calculate the z-component of the vorticity and the horizontal 
divergence as, 

\begin{equation}
 \centering
  (\bm{\nabla} \times \bm{v})_{z}=\bigg(\frac{\partial v_{y}}{\partial x} - \frac{\partial v_{x}}{\partial y}\bigg),
 \label{curlvz}
\end{equation}

\begin{equation}
 \centering
 (\bm{\nabla}.\bm{v})_{h}=\bigg(\frac{\partial v_{x}}{\partial x} + \frac{\partial v_{y}}{\partial y}\bigg) .
 \label{divvh}
\end{equation}
%where, $v_{x}$, $v_{y}$ are the horizontal velocity components. 
Calculation of kinetic helicity $H_{k}$ requires the vertical component of $\bm{v}$ and its gradient, which we 
do not have. We follow \citep{Rudiger1999} in deriving a proxy for kinetic helicity from the calculated vertical 
component of vorticity and horizontal divergence, 
\begin{equation}
 \centering
H_{k,proxy}=\frac{\langle (\bm{\nabla}.\bm{v})_{h})(\bm{\nabla} \times \bm{v})_{z}) \rangle}{\langle (\bm{\nabla}.\bm{v})_{h})^{2}\rangle^{1/2} \langle (\bm{\nabla} \times \bm{v})_{z})^{2} \rangle^{1/2}}.
\end{equation}
%{\bf The $H_{k}$ is found to be weak in the theoretical calculations, hence \citet{Brandenburg1995} derived relative 
This proxy for kinetic helicity is similar to the relative kinetic helicity, $H_{k,rel}$, used by 
\citet{Brandenburg1995} 
\begin{equation}
 \centering
H_{k,rel}=\frac{\langle \bm{\omega} .\bm{v}\rangle}{\langle \omega ^{2}\rangle ^{1/2}\langle v ^{2}\rangle ^{1/2}}.
\end{equation}
in situations dominated by two-dimensional flows.

\section{Results: vortical motions, kinetic helicity and the magnetic field}
Spatial maps of horizontal divergence, $d_{h} = (\bm{\nabla}.\bm{v})_{h}$, the vertical vorticity 
$\omega_{z} = (\bm{\nabla} \times \bm{v})_{z}$, and the kinetic helicity $H_{k}$ (hereafter we denote 
$H_{k,proxy}$ defined above simply as $H_{k}$), derived at each time step as described in the previous Section, 
form our basic fluid dynamical quantities. To improve the signal-to-noise of these measurements, we use a running 
temporal average over about 4.5 minutes, {\it i.e.} average of six individual measurements; this is found suitable 
as typical life-time of granules is about 5 - 7 minutes. This running average is taken on the flows derived
but not on the cross-correlations of LCT to avoid missing any granular signals that have life-time smaller or close to the
averaging interval.
An example map of full 14 hr temporal average of 
$\mathbf d_{h}$ and $\omega_{z}$  with overlaid contours of similarly averaged $B_{\rm{LOS}}$ is shown in 
Figure \ref{contourplot}.

The velocity maps $v_{x}$ and $v_{y}$ calculated from Doppler observations show a systematic variation in x- and 
y-directions, and the magnitude of change across the spatial extent covered is about $0.4 ms^{-1}$. This is attributed 
to the 'shrinking Sun effect' \citep{Lisle2004, Langfellner2015}  that shows an apparent disk-centered (or 
radially directed) inflows. Origin of this constant 
flow signal (i.e. time independent) is not fully known, although it has been attributed to selection bias of 
LCT method  and to insufficient resolution of the instrument to resolve fully the granules on the solar 
surface \citep{Lisle2004}. Whatever the origin, this constant disk-centered flow signal is
easily determined by taking temporal averages (of both $v_{x}$ and $v_{y}$), spatillay smoothing and 
obtaining a low degree 2-D fit of it.So determined background artefact is then removed by subtracting it out 
from maps $v_{x}$ and $v_{y}$ at every time step.

In this study, we examine (1) the hemispherical dependence of the signs of $\omega_{z}$ or $H_{k}$ arising from the 
Coriolis force, 
(2) how the magnetic field modifies the relationship between $d_{h}$ and $\omega_{z}$ or $H_{k}$, and 
(3) how these quantities themselves scale against the strength of the magnetic field. Since these quantities 
are highly dynamic with typical time scales of the order of granular life-time, we derive the relationship 
between these quantities at each time step of measurement.
This we do by determining, at each time-step, the
dependence of $\omega_{z}$ (or $H_{k}$) on $d_{h}$ and $B$ by calculating their average values over chosen small 
intervals (bin sizes) in $B$ and $d_{h}$: 10 G \footnote{ We note that the use of $B = B_{LOS}$
leads to, when much of the magnetic field is vertically oriented on the surface, a systematic bias towards
lower field strengths that are about $cos(\theta)$ times the true value for an angular distance of $\theta$\textdegree~
from the disk center. Since the maximum off-center location does not exceed $\theta = 30$\textdegree, we have
at the most a 14\textdiscount~ lower values for $B$. However, average deviations of derived dependences on $B = B_{LOS}$
would be off from true values of $B$ by a much smaller amount than 14\textdiscount.} bins in $B$ (a magnetic bin) and 
20$\ \mu s^{-1}$ bins in $d_{h}$. This is implemented by sub-dividing every magnetic bin, {\it i.e.} pixels having 
magnetic field spread of 10 G, in terms of $d_{h}$ with a bin size 20$\ \mu s^{-1}$. 
We do this for both signed and absolute values of $\omega_{z}$ or $H_{k}$, denoting them respectively as 
$\omega_{z}(d_{h},B)$ or $H_{k}(d_{h},B)$ and $\omega_{z}^{ab}(d_{h},B)$ or $H_{k}^{ab}(d_{h},B)$; in the former 
case of averages of signed values we perform the calculations for northern and southern hemispheric regions 
separately so as to check the hemispheric trends such as those introduced by Coriolis force, and in the latter 
case of average of absolute values we combine both hemispheric regions together.
The fill-factor for magnetic field is small ({\it i.e.}, much of the area is occupied by $|B| \approx $ 0 G pixels, 
refer to Fig. \ref{contourplot}) and, in general, area occupied by magnetic pixels goes down sharply as $|B|$ increases (Refer
to the bottom panel of Figure 3, where a histogram of area (in logarithmic scale) against $B$ is plotted). In 
such situations a better statistic is provided by median values rather than averages. We tested this by taking
median values over the chosen bins of $B$ and $d_{h}$ and found, however, that the derived relationships are nearly 
the same for averages and medians. 

\begin{figure*}[ht]
\epsscale{1.05}
\centering
\plotone{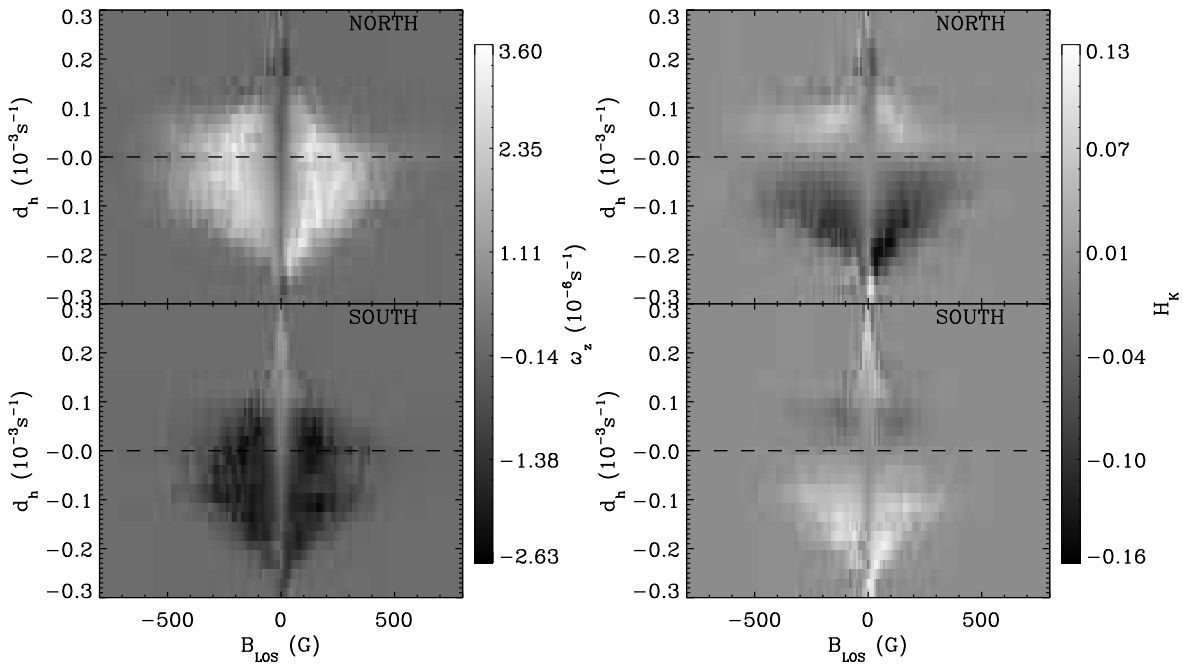}
\caption{\footnotesize Vertical (z-component) vorticity, $\omega_{z}(d_{h},B)$ (left panels), and kinetic 
helicity, $H_{k}(d_{h},B)$ (right panels), binned against LOS magnetic field, $B$ (x-axis), and horizontal divergence, $d_{h}$ (y-axis).
The results here are calculated from local correlations tracking of granular motions imaged in HMI continuum intensities, and are averages 
over quiet-network in nineteen large regions covering northern and southern hemispheres. See text for further details.}
\label{2dplotIc}
\end{figure*}

\begin{figure*}[ht]
\epsscale{1.05}
\centering
\plotone{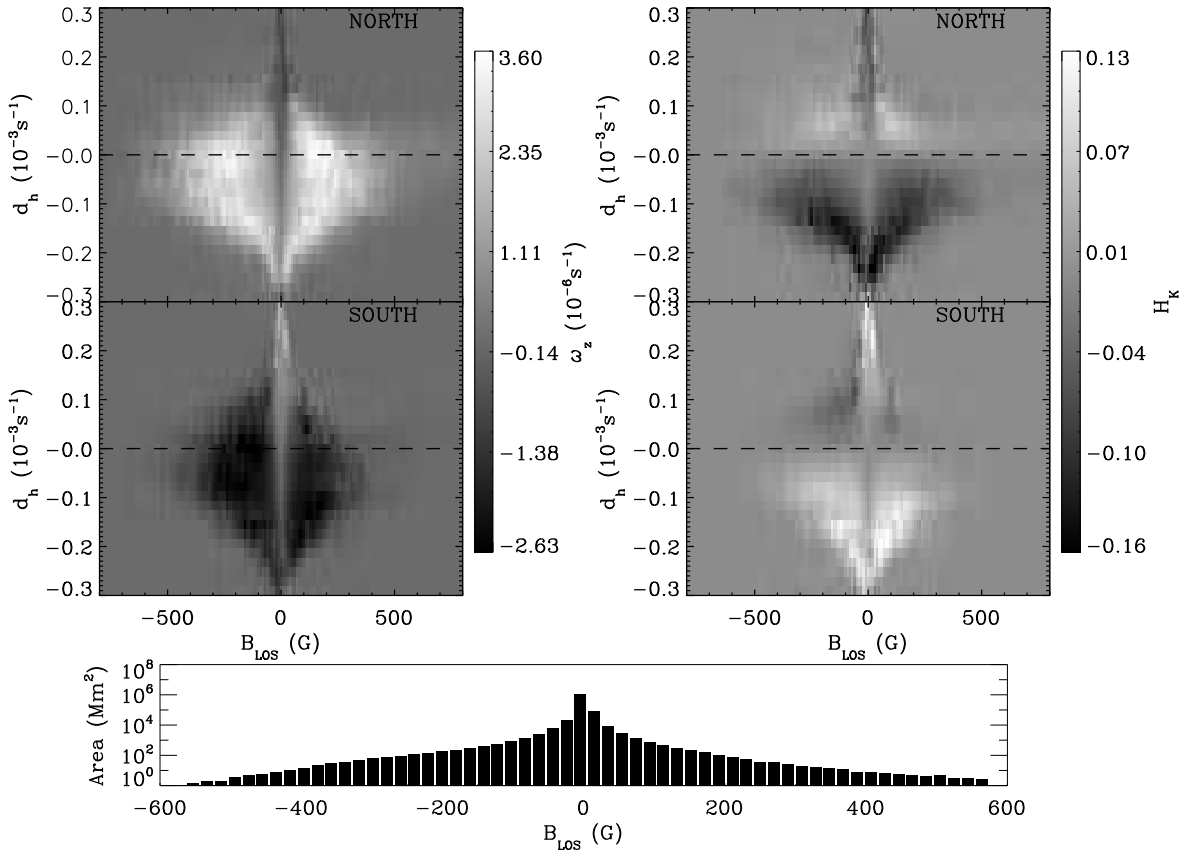}
\caption{\footnotesize The $\omega_{z}(d_{h},B)$ and $H_{k}(d_{h},B)$ are the same as plotted in Fig. (\ref{2dplotIc}) but 
are derived from correlation tracking of granular motions imaged in HMI Doppler velocities. The bottom panel shows a histogram
of area (in logarithmic scale) occupied by magnetic pixels in units of $Mm^{2}$ against $B=B_{LOS}$.}
\label{2dplotV}
\end{figure*}

The above analysis process is repeated for each region, and average $\omega_{z}(d_{h},B)$, $\omega_{z}^{ab}(d_{h},B)$, 
$H_{k}(d_{h},B)$ and $H_{k}^{ab}(d_{h},B)$ are determined from those for all the regions, making sure that the 
included areas contain only the quiet-Sun magnetic network. The resulting average relationships 
$\omega_{z}(d_{h},B)$ and $H_{k}(d_{h},B)$ determined for northern and southern hemispheric regions separately 
are shown in Figures \ref{2dplotIc} and \ref{2dplotV}: results based on LCT velocities derived using HMI 
continuum intensities $I_{\rm{c}}$ are in Figure \ref{2dplotIc} and that from Doppler velocities are in 
Figure \ref{2dplotV}, and they agree well. It is seen that the Doppler velocities yield a little less noisy 
results for $\omega_{z}(d_{h},B)$ and $H_{k}(d_{h},B)$ and hence we use these for further analyses in the 
following Sections. It is to be noted that the signed averages $\omega_{z}(d_{h},B)$ ($H_{k}(d_{h},B)$) 
have cancellations and hence measure only the excess of one sign of rotation [either clock-wise (negative) or 
counter-clockwise (positive) rotation] over the other. Hence, vanishing of these averages do not necessarily mean 
absence of rotations, and this is easily checked by the averages of absolute values 
$\omega_{z}^{ab}(d_{h},B)$ ($H_{k}^{ab}(d_{h},B)$) shown in Figures \ref{cavsdivaV} and \ref{kavsdivaV}. As can 
clearly be seen, the excesses $\omega_{z}(d_{h},B)$ ($H_{k}(d_{h},B)$) of one sign over the other are about 
one-tenth of $\omega_{z}^{ab}(d_{h},B)$ ($H_{k}^{ab}(d_{h},B)$). Further, we see that such cancellations are 
the largest for non-magnetic (quiet) flows. The dominance of one sign of $\omega_{z}(d_{h},B)$ for flows around 
magnetised regions is due to the phenomenon of flux expulsion \citep{Proctor1982} that leads to magnetic flux occupying 
preferentially the inflow (negative $d_{h}$) locations. We discuss this aspect further later in the next Section.

\subsection{Vorticity - divergence correlation: effects of Coriolis force}
Results in Figures \ref{2dplotIc} and \ref{2dplotV} show that, to a large extent, the sign of $\omega_{z}$ ($H_{k}$) 
is positive (negative) or counter-clockwise rotation in the northern hemisphere and negative (positive) or clockwise 
rotation in the southern hemisphere. It is to be noted that, in the absence of Coriolis and any other large scale 
force, the flows are expected to have a roughly equal distribution of clock-wise (negative) and anti-clockwise 
rotations, and hence a signed average of $\omega_{z}$ or $H_{k}$ over magnetic and divergence bins is expected to 
yield near-zero values due to cancellation among positive and negative vorticities. However, as earlier studies have 
shown, supergranular scale flows are indeed subject to Coriolis force, and as results in Figures \ref{2dplotIc} 
and \ref{2dplotV} show even the smaller-sized inflows (negative $d_{h}$) show predominant hemispheric sign 
pattern consistent with the Coriolis effect. A closer look at this striking pattern requires understanding 
first the $d_{h} - \omega_{z}$ relationship for the non-magnetic flows. 

\begin{figure*}[ht]
\centering
\plotone{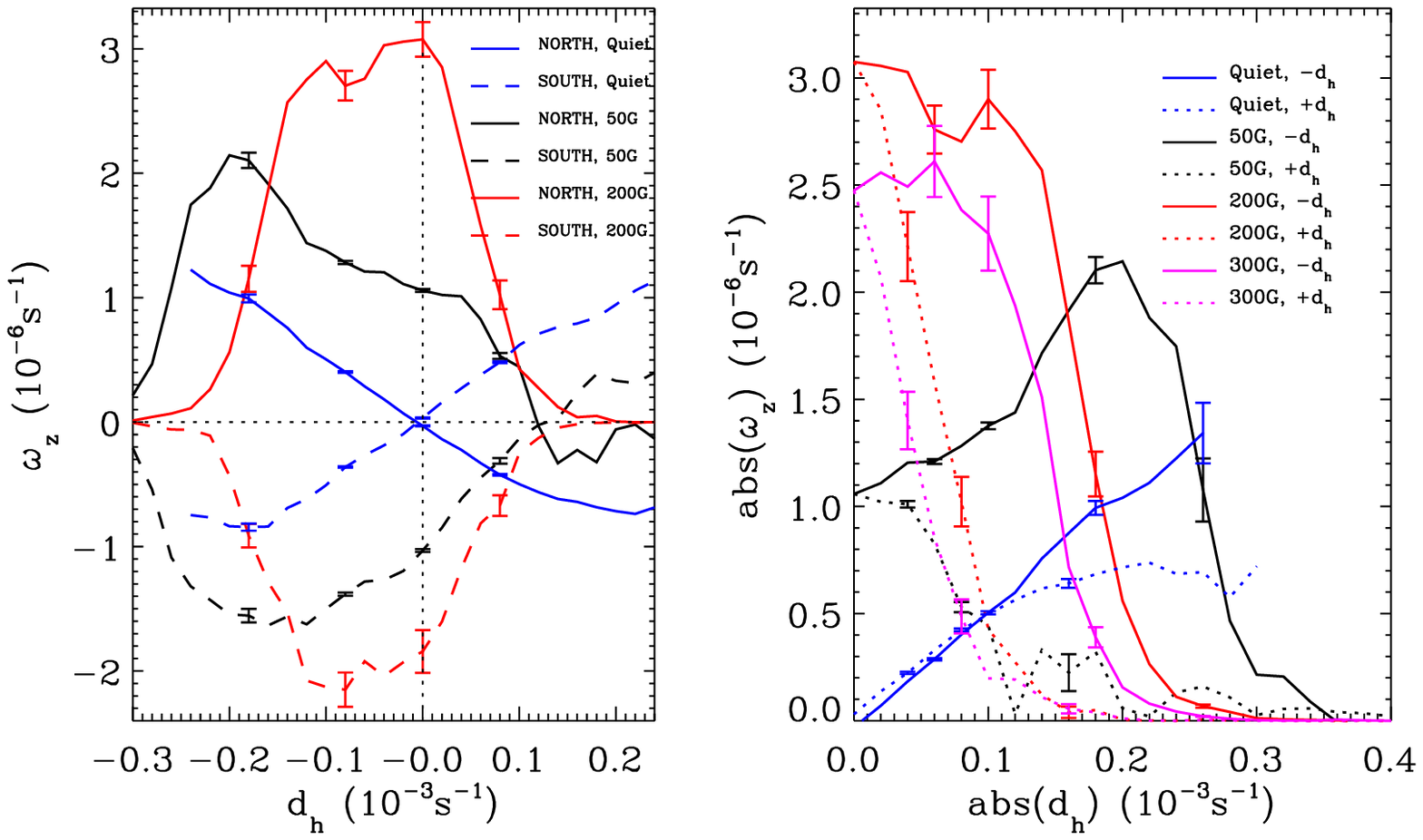}
\caption{\footnotesize Averages over different field strength ranges, as marked in the panels,  
of vertical vorticities $\omega_{z}(d_{h},B)$ calculated from results shown in Figure \ref{2dplotV}. 
The left panel shows signed $\omega_{z}$ against signed $d_{h}$, whereas the right panel shows the magnitudes of 
these quantities for an easy comparison of inflow (solid curves) and outflow (dotted curves) regions. 
The 300 G curves added in the right panel are to show that the vorticity values decrease 
beyond 200G. They are not shown in the left panel for the sake of clarity as we include both northern and 
sourthern hemisphere results in the same plot. For the 
magnitudes in the right panel, we have used those only for the north region. Refer to the text for further details.}
\label{cavsdivsV}
\end{figure*}

\begin{figure*}[ht]
\centering
\plotone{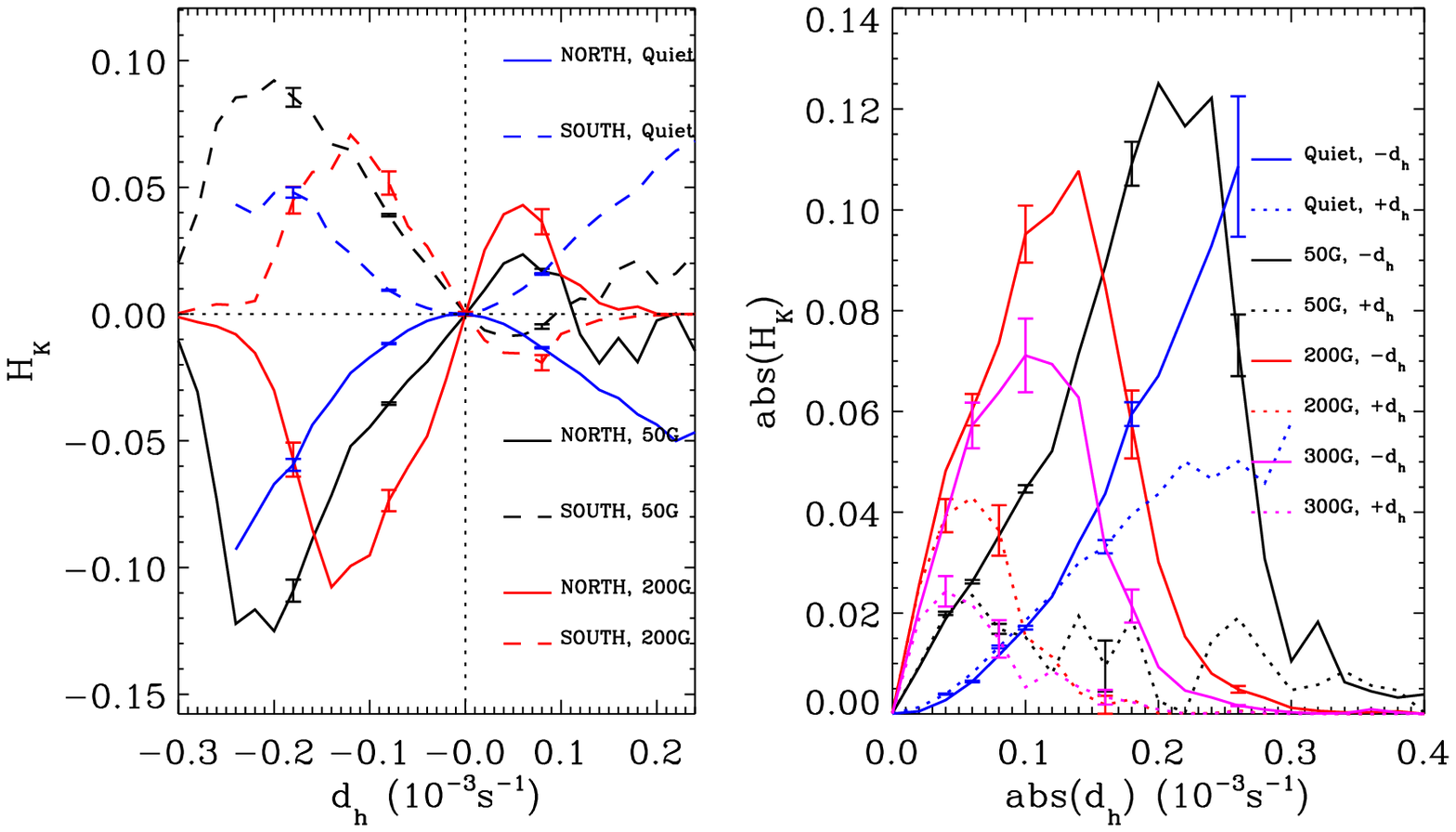}
 \caption{\footnotesize Similar plot as Figure \ref{cavsdivsV} but for signed averages of kinetic helicity, $H_{k}(d_{h},B)$.}
 \label{kavsdivsV}
\end{figure*}
Quiet-Sun areas devoid of significant magnetic field is captured in the central vertical area close to $|B| = 0$G in 
Figures \ref{2dplotIc} and \ref{2dplotV}. We take the average of $\omega_{z}$ within $|B| < \bm{15}$G, {\it i.e.} 
a horizontal average over three magnetic bins of -10, 0, and 10 G in Figure \ref{2dplotV}, to determine 
$\omega_{z}^{q}(d_{h})$ = $\omega_{z}(d_{h},B=0)$ for the non-magnetic flows. It is to be noted that the 
observational errors in HMI LOS magnetograms are about 10 G \citep{Scherrer2012} and hence the pixels within 
 $|B| < \bm{15}$G can be considered as non-magnetic. The resulting variation of $\omega_{z}^{q}(d_{h})$ is shown as 
blue lines in Figure \ref{cavsdivsV} (other colored lines in this figure show averages over different ranges of 
$|B|$, and we discuss them in the next Section). Signed values of $\omega_{z}$ against $d_{h}$ are shown in the 
left panel of Figure \ref{cavsdivsV}, and a comparison of absolute amplitudes of $\omega_{z}$ at inflows 
and outflows is shown in the right panel by plotting them against absolute $d_{h}$. Results plotted similary 
for $H_{k}$ are in Figure \ref{kavsdivsV}. In these figures, we show only the results obtained from LCT 
velocities derived using HMI Doppler velocities (results obtained from HMI $I_{\rm{c}}$ are very similar). 
The variation of $\omega_{z}^{q}(d_{h})$ clearly brings out the effect of Coriolis force on fluid flows on 
the supergranular scale, {\it viz.} a radial outflow (positive $d_{h}$) rotates clock-wise (negative $\omega_{z}$) 
and a inflow (negative $d_{h}$) rotates counter-clock-wise (positive $\omega_{z}$) in the northern hemisphere 
and vice versa in the southern hemisphere. These results are in agreement with earlier known rotation properties 
of supergranular flows \citep{Duvall2000,Gizon2003}. Further the magnitudes of $\omega_{z}^{q}$ derived as above 
increase more or less linearly against magnitudes of $d_{h}$. 

\begin{figure*}[ht]
\centering
\plotone{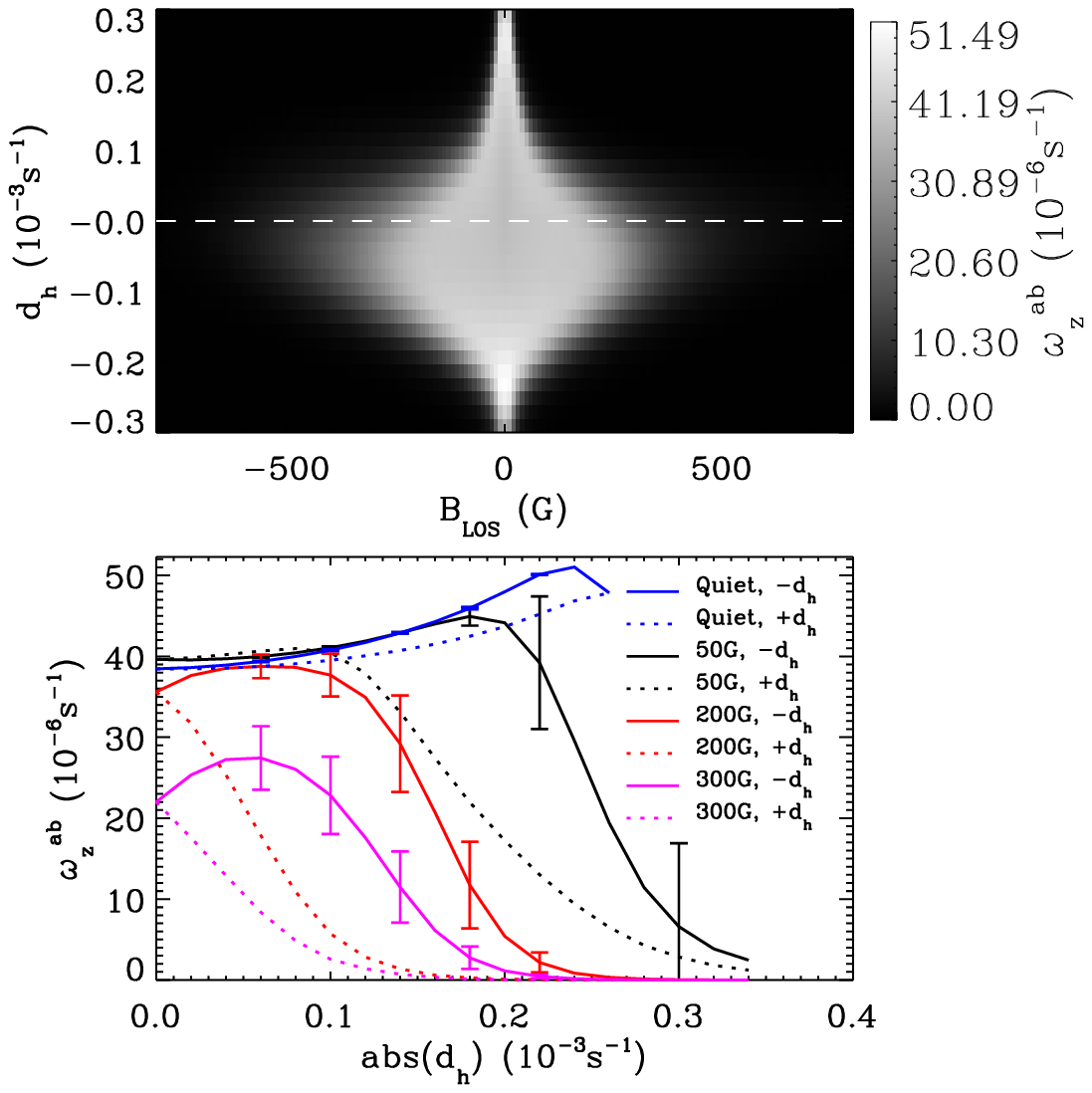}
\caption{\footnotesize Absolute values of vertical (z-component) vorticity, $\omega_{z}^{ab}(d_{h},B)$ (top panel),
binned against LOS magnetic field, $B$ (x-axis), and horizontal divergence, $d_{h}$ (y-axis). Bottom panel shows the cross-sections
averaged over different field strength ranges the same as in the right panel of Figure \ref{cavsdivsV}. See text for further details. }
\label{cavsdivaV}
\end{figure*}
\begin{figure*}[ht]
\centering
\plotone{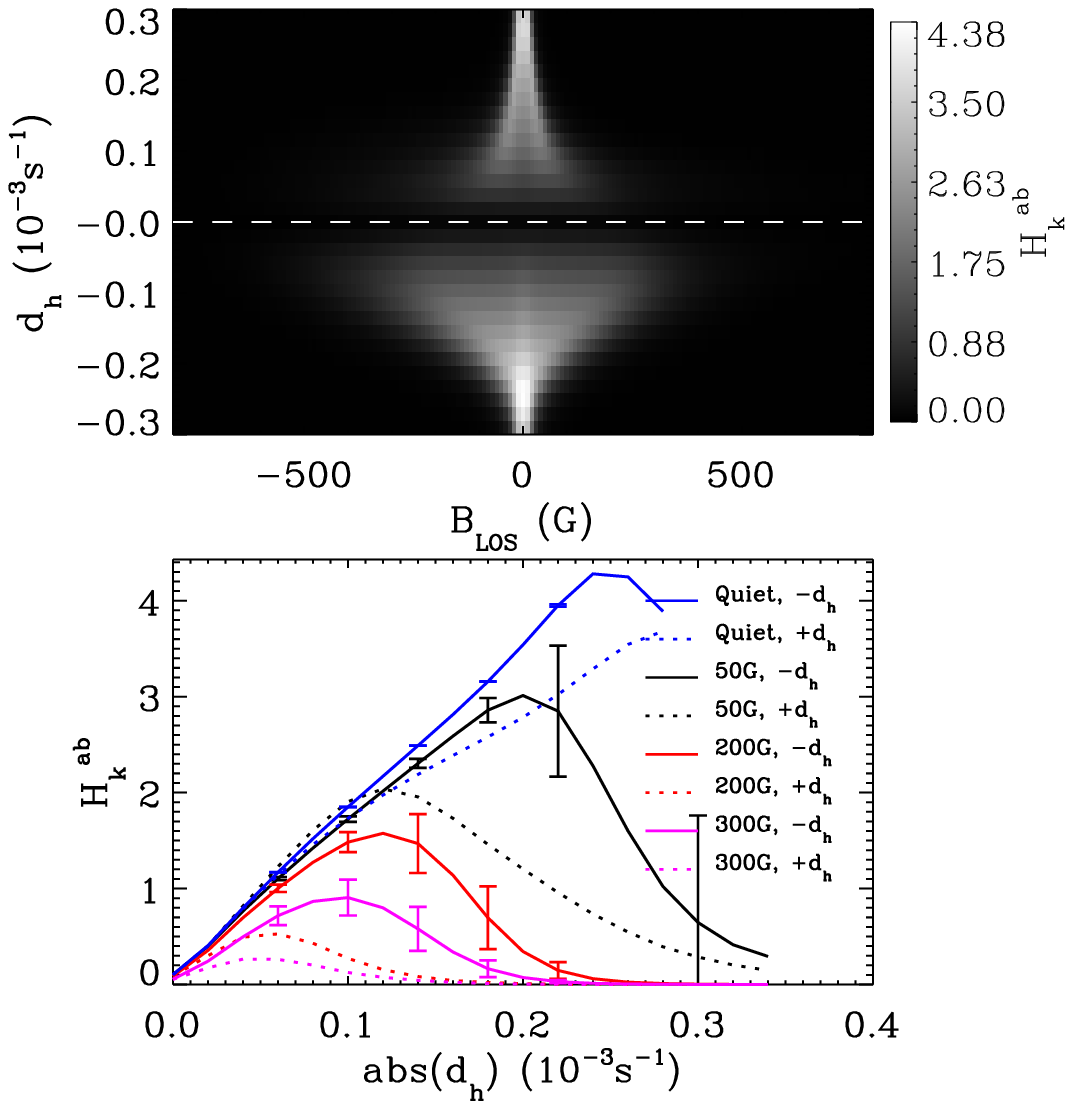}
 \caption{\footnotesize Similar plot as Figure \ref{cavsdivaV} but for averages of absolute values of kinetic helicity, $H_{k}^{ab}(d_{h},B)$.}
 \label{kavsdivaV}
\end{figure*}

Although the above result indicates that the predominant variation 
of $\omega_{z}^{q}(d_{h})$ in quiet-Sun flows at the scales that we are measuring is due to Coriolis effect, 
it is expected that not all size vortex flows have the signs following the Coriolis effect. Especially, any smaller 
scale vortex flows at granular inflows or junctions within the supergranular cells can potentially be of 
larger magnitude and not influenced by the Coriolis force. This indeed turns out to be true as the averages of 
absolute values, $\omega_{z}^{ab}(d_{h},B)$ ($H_{k}^{ab}(d_{h},B)$), shown in Figures \ref{cavsdivaV} and 
\ref{kavsdivaV} portray. The top panels in these figures show the 2-D binned maps, while the bottom panels show 
the non-magnetic $\omega_{z}^{ab,q}(d_{h})$ (blue lines) and averages over a few selected magnetic field ranges 
(same as done in Figures \ref{cavsdivsV} and \ref{kavsdivsV}) against abs($d_{h}$) for a comparison of outflows and 
inflows. In the analysis and plots here, the error bars represent standard deviations within the ranges of 
magnetic bins used. Firstly, as noted earlier, magnitudes of vorticities (determined through averages of 
absolute values) are about ten times those of the excesses determined to be resulting from the 
hemispherical preference of one sign over the other due to the Coriolis force (comapre the values in 
Figures \ref{cavsdivsV} and \ref{kavsdivsV} with those in Figures \ref{cavsdivaV} and \ref{kavsdivaV}). 
Magnitudes of vorticities estimated in our work here compare very well with earlier local helioseismic results from 
\citet{Gizon2003, Duvall2000} and with local helioseismic as well as LCT analyses of \citet{Langfellner2015}.
Secondly, in agreement with the earlier findings \citep{Wang1995}, the inflow regions have slightly excess 
vorticity (and helicity) as compared to outflow regions for non-magnetic flows (blue curves in Figures 
\ref{cavsdivaV} and \ref{kavsdivaV}). We discuss the magnetic modifications in the following Section. 

\subsection{Magnetic effects: transfer and redistribution of vorticity}
To study how the above described non-magnetic relationship $\omega_{z}^{q}(d_{h})$ is influenced by the magnetic 
field, we average $\omega_{z}(d_{h},B)$ and $\omega_{z}^{ab}(d_{h},B)$ (as well as $H_{k}(d_{h},B)$ and 
$H_{k}^{ab}(d_{h},B)$) over three different ranges of magnetic field: 20 - 80 G (denoted as 50 G), 150 - 250 G
(denoted as 200 G), and 250 - 350 G (denoted as 300 G) centered around 50, 200 and 300 G. 
The results are shown in the Figures \ref{cavsdivsV}, \ref{kavsdivsV}, \ref{cavsdivaV} and \ref{kavsdivaV}. Firstly, 
we see from Figures \ref{2dplotIc} and \ref{2dplotV} that the sign of $\omega_{z}(d_{h},B)$ over magnetized regions 
does not follow the dependence on $d_{h}$ expected from the action of Coriolis force: while the non-magnetic 
($<\pm$10G) central vertical regions of Figures \ref{2dplotIc} and \ref{2dplotV} show the sign change (the blue 
curves in the left panel of Figure \ref{cavsdivsV}) through $d_{h}$ = 0., the magnetised regions show 
predominantly one sign, {\it i.e.} positive in the north and negative in the south, which should hold only for 
inflows or converging flows (negative $d_{h}$) if the Coriolis effect is the cause. However, magnetic fields over 
outflows or diverging flows (positive $d_{h}$) too show the same sign as that seen over converging (negative 
$d_{h}$) regions, although it is noted that the majority of the magnetic fluxes lie over the coverging or 
inflow (negative $d_{h}$) regions. It is clear that the dominantly inflow located magnetic 
fields are connected to those located in the interiors of supergranules (with positive $d_{h}$), and 
that such connectivity transfers vorticity from the inflow (negative $d_{h}$) 
regions to fluid surrounding such connected fields within the supergranules (positive $d_{h}$). This magnetic 
connectivity is expected in the predominantly mixed-polarity flux, which can have a large horizontal component or 
loops arching above the photosphere. This transfer can happen in either direction, but it is seen in 
Figures \ref{2dplotIc} and \ref{2dplotV} that tranfer of vorticity (and kinetic helicity) is predominantly from 
inflow regions (negative $d_{h}$) to outflows or diverging flows (positive $d_{h}$); 
this is consistent with the fact that stronger magnetic fields with larger fluxes are found mainly over the inflow regions. The cross-sections over 
three different magnetic field ranges plotted in Figures \ref{cavsdivsV}, \ref{kavsdivsV}, \ref{cavsdivaV} and 
\ref{kavsdivaV} show the above described signatures of transfer of vorticity from 
inflow to outflow regions more clearly, and also show how the strengths of the magnetic field influence 
this phenomenon.

We add a caveat here to the above inferences on possible transfer of vorticity from supergranular inflow (negative $d_{h}$)
to outflow (positive $d_{h}$) regions: the rearrangement of magnetic flux during the process of 
evolution of supergranules, viz. ``dying" inflow that is replaced by the outflow of a new supergranule, can lead to our analyses 
finding vorticity in outflow regions that matches the vorticity in inflow regions (as described above). This can arise
because the newly formed outflow (in the place of old inflow) has not had time to sweep the flux to the new inlfow region and
the Coriolis force too has not had time to act and hence reverse the vorticity. However, we note that the 
averaging over 14 hours, which is more than about half the life time of a supergranule and hence is long enough for both
the above processes to establish, should tend to smooth out such signatures of dying and newly forming supergranular flows. 
Nevertheless, a residual signature of this process is certainly possible in our analyses.

In addition, we find that these differences between inflow and outflow vorticities (and helicities) are much more 
pronounced for flows around magnetic fields -- refer to the black (50 G), red (200 G) and pink (300 G) curves in 
Figures \ref{cavsdivaV} and \ref{kavsdivaV}. It is to be noted however that these enhanced excesses of inflow 
vorticities, both in signed averages (Figures \ref{cavsdivsV} and \ref{kavsdivsV}) as well as in averages of 
absolute values (Figures \ref{cavsdivaV} and \ref{kavsdivaV}), also depend on the fact that 
much of the magnetic fields, especially stronger concentrations, are located in inflow regions. This is clearly 
seen in the 2D maps in Figures \ref{2dplotIc} and \ref{2dplotV} and in the top panels of Figures \ref{cavsdivaV} and 
\ref{kavsdivaV}, which show the large asymmetry in the areas occupied by non-zero vorticities in the 
$B - d_{h}$ region: relatively weaker fields are found in the supergranular outflows (positive $d_{h}$), 
and majority of the magnetic flux and stronger field concentrations are in the inflow (negative $d_{h}$) 
regions. A basic feature in all the above magnetic effects is that the linear relationship characterising 
$\omega_{z}^{q}(d_{h})$ has given way to more complicated variations. We present and describe the dependences on
magnetic field strength further in the following Section.

\subsubsection{Magnetic suppression of fluid vorticity and helicity}
\begin{figure*}[ht]
\centering
\plotone{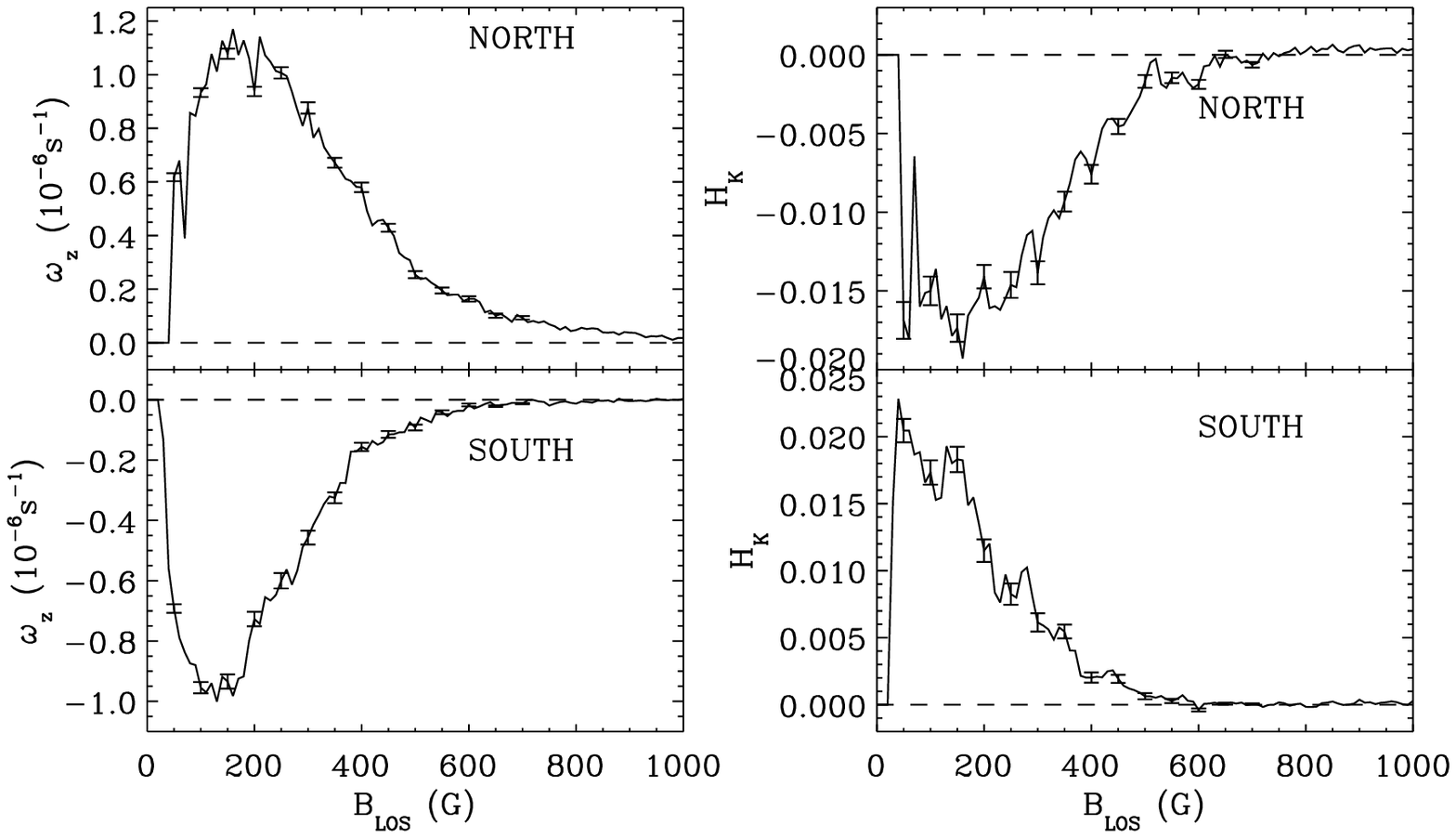}
\caption{\footnotesize The left panels show vertical vorticities, $\omega_{z}(B)$, that are averaged over divergence
bins in Figure \ref{2dplotV}, plotted against field strength $|B|$ for northern and southern hemispheres separately. In the right panels
are the similarly averaged kinetic helicities $H_{k}(B)$. The error bars represent standard errors estimated from individual
measurements. Horizontal dashed lines in the panels mark the zero level of the ordinates.}
\label{ca1vsBV}
\end{figure*}

\begin{figure*}[ht]
\centering
\plotone{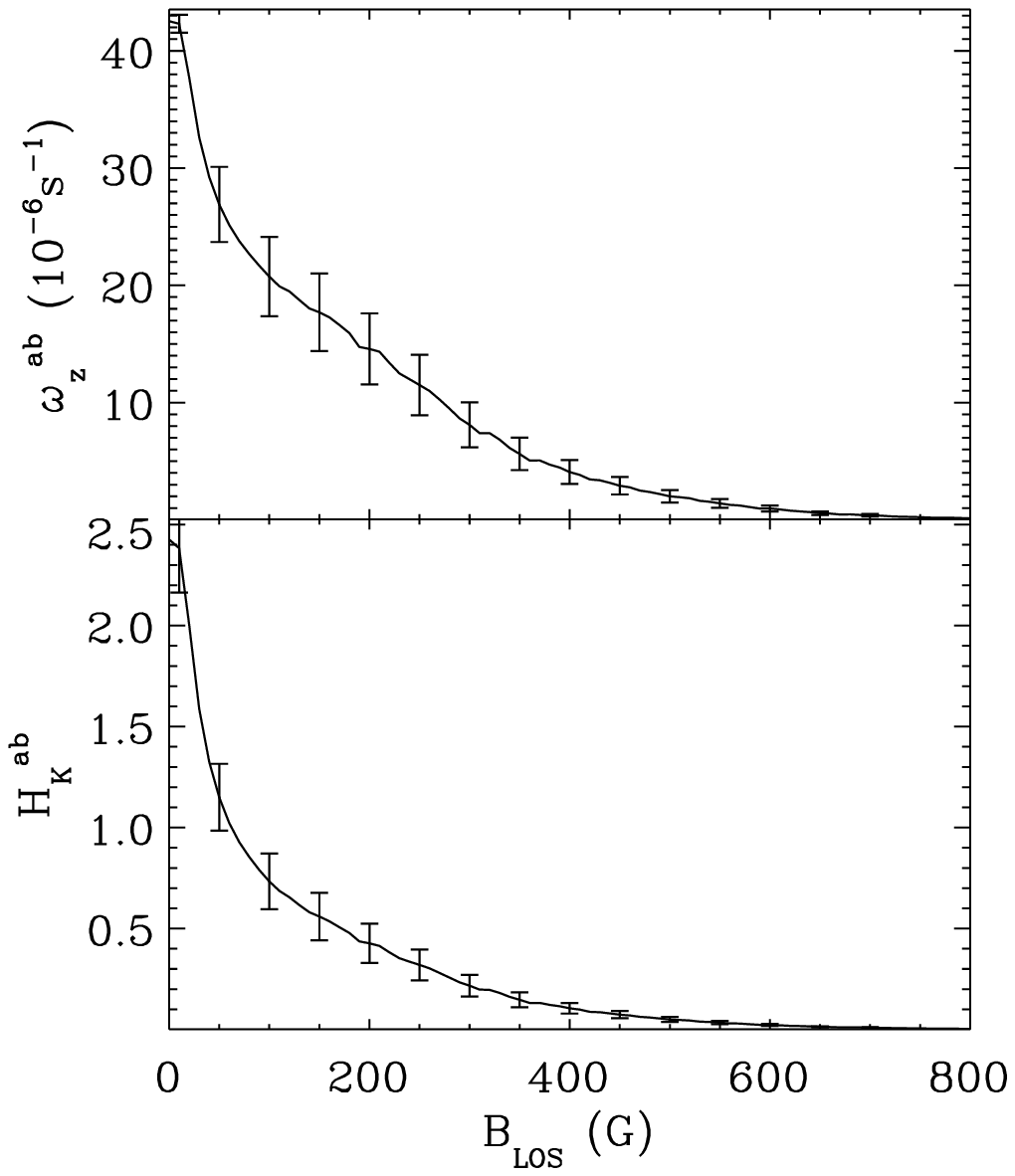}
\caption{\footnotesize The same as in Figure \ref{ca1vsBV}, but for averages of absolute vorticties, $\omega_{z}^{ab}(B)$ (top panel), and
kinetic helicities $H_{k}^{ab}(B)$ (bottom panel). Here, the north and south regions have been averaged together.}
\label{cavsBV}
\end{figure*}

To examine the influence of magnetic field strengths on the sign and amplitudes of $\omega_{z}$ and $H_{k}$ 
(Figure \ref{2dplotV}), and on the absolute magnitudes of $\omega_{z}^{ab}$, and $H_{k}^{ab}$ (top panels of 
Figures \ref{cavsdivaV} and \ref{kavsdivaV}), we take averages of these quantities over the divergence bins. 
These results are shown in Figures \ref{ca1vsBV} and \ref{cavsBV}. As noted earlier, at $|B|$=0 G, the opposite 
contributions to $\omega_{z}$ and $H_{k}$ from the positive (outflow) and negative (inflow) $d_{h}$ regions 
cancel out yielding near-zero values for these quantities. However, averages of absolute values $\omega_{z}^{ab}$ 
and $H_{k}^{ab}$ show that the largest magnitude vorticities are to be found over non-magnetic flows, and they 
decline sharply as $|B|$ increases. The excesses $\omega_{z}(B)$ and $H_{k}(B)$, however, show a rapid increase 
at low magnetic field strengths, reaching a maximum between about 150 and 200 G. The sign pattern of magnetic 
field correlated $\omega_{z}$ and $H_{k}$ preserves that expected from Coriolis force 
action on inflows (negative $d_{h}$ regions). Taken together, these results show that a major contribution 
to larger magnitudes of vorticities and kinetic helicities in inflow regions, as compared to those over 
diverging flows in the interiors of supergranules, leading to the hemispheric excesses seen in Figure \ref{ca1vsBV} 
are due to the predominance of magnetic fields at inflow locations. It is also possible that the supergranular 
inflows have more vigorous swirls around them due to larger thermal perturbations that strong magnetic flux 
concentrations cause in the solar photospheric layers: cooler magnetic field trapped photospheric gas acts as 
channels for further heat (radiative) loss vertically outwards accelerating the flows surrounding hotter 
plasma towards them, thus enabling stronger inflows and vorticities.

As magnetic field increases further beyond equipartition strength, which is typically about 300 - 400 G for the 
solar photosphere, the back-reaction of the magnetic field via Lorentz force, involving both the pressure and 
tension forces, acts to inhibit the flows. This effect eventually becomes dominant and acts to compensate the 
contributions from the Coriolis force assisted combinations of inflows and magnetic fields leading to the 
decreases seeen in Figure \ref{ca1vsBV}. As to the averages of absolute magnitudes $\omega_{z}^{ab}$ and $H_{k}^{ab}$ 
shown in Figure \ref{cavsBV}, it is clear that, when we ignore the outflow - inflow asymmetry in the magnetic field 
locations and the resulting influence of Coriolis force, the net effect of magnetic forces is to suppress vortical 
motions. There might also be contributions due to the fact that field strengths $B$ here (from HMI) are flux 
densities within the resolution element of the instrument, and hence increase of $|B|$ here may mainly reflect 
the increase of flux while the actual field strengths being already in strong saturated super-equipartition values. 
That the photospheric small-scale network field concentrations outlining supergranular inflows are predominantly 
of super-equipartition kG field strengths is a well established fact. Hence, the results in Figure \ref{cavsBV} for 
vorticities and kinetic helicities may correspond to the so-called highly $\alpha$-quenched state, if photospheric 
flows are considered for a (helical)-turbulent local dynamo (refer to, {\it e.g.}, \citet{Brandenburg2005} and 
references therein). In non-linear mean-field dynamo theory, both the hydrodynamic ($\alpha_{k}$) and magnetic 
($\alpha_{m}$) parts of the the $\alpha$-effect are coupled via the magnetic field, and explicit algebraic 
expressions for the dependences $\alpha_{k}(B)$ and $\alpha_{m}(B)$ are available 
\citep{Rogachevskii2000, ZhangH2006}. The $\alpha_{k}$ is proportional to $H_{k}^{ab}$ shown in Figure \ref{cavsBV}. 
However, we note that theoretical quenching functions that determine $\alpha_{k}(B)$ are in terms of mean 
magnetic field $B$ that develops out of dynamo action at the small-scales, whereas in our observations and 
analysis here $B$ refers to the field at the small-scale. A more detailed look at $H_{k}^{ab}(B)$ or $\alpha_{k}(B)$, 
their spectra, both spatial and temporal, and their relations to various other local and global properties of 
solar magnetic fields in the context of dynamo mechanisms is beyond the scope of work reported here. 
Moreover, higher spatial resolution to resolve sub-granular scale flows and magnetic fields than provided by 
HMI is important to address such details.

\section{Summary and Discussion}

We have derived and analysed horizontal fluid motions on the solar surface over large areas covering the 
quiet-Sun magnetic network using local correlation tracking of convective granules imaged in continuum 
intensity and Doppler velocity by the HMI onboard SDO. We have studied the relationships between fluid 
divergence and vorticity, and that between vorticity (kinetic helicity) and magnetic field. These relationships 
are derived through both signed and absolute averages of $\omega_{z}$ and $H_{k}$ over time and space. 
These latter dependences are studied through fluid divergence $d_{h}(x,y,t)$ and magnetic field $B(x,y,t)$, 
yielding $\omega_{z}(d_{h},B)$, $H_{k}(d_{h},B)$, $\omega_{z}^{ab}(d_{h},B)$, and $H_{k}^{ab}(d_{h},B)$ 
(Figures \ref{2dplotIc}, \ref{2dplotV}, \ref{cavsdivsV}, and \ref{kavsdivsV}). The main results obtained can be 
summarised as below:\\
(1) The correlations between vorticity and divergence of non-magnetic flows at the supergranular scale have 
the dominant hemispheric sign pattern brought about by the action of the Coriolis force. The rotations of 
outflows and inflows are roughly of equal magnitude and of opposite sign, although there is a slight excess of 
inflow vorticities which increase as magnitude of divergence increases. Further, the non-magnetic vorticities 
scale linearly with the divergences.\\
%In absolute magnitudes, non-magnetic flows show the largest values for vorticities and kinetic helicities.\\
(2) For magnetised flows, the sign pattern corresponding to the inflows dominate over the whole divergence field, 
{\it i.e.} even the outflow regions (positive divergence, interiors of supergranules) exhibit rotations expected 
of the action of Coriolis force acting on inflows. In other words, for magnetised flows the dominant sign 
pattern is negative helicity $H_{k}$ (positive $\omega_{z}$) in the north and positive helicity $H_{k}$ 
(negative $\omega_{z}$) in the south. We have identified and attributed this to the mechanism of transfer of 
vorticity from network inflow regions to interiors of supergranules by the magnetic connectivity or by 
horizontal component of magnetic field, with a caveat that supergranular evolution (``dying" inflow that is replaced by 
the outflow of a new supergranule) too can contribute to the observed signatures.\\
(3) The excess of inflow vorticities over those of outflows increases dramatically for magnetised flows. 
This has been identified as due to the preferential inflow locations of magnetic fields resulting from the 
convective flux expulsion mechanism. Contributions of instrinsically larger rotations of inflows around 
stronger fields there due to thermal causes, however, can also play a role in the above asymmetries between 
outflow and inflow rotations.\\
(4) In terms of absolute magnitudes of vorticities, it has been found that the non-magnetic flows have the 
largest values. As a function of magnetic field strengths, as observed by HMI, magnitudes of absolute vorticities 
decrease almost exponentially bringing out the magnetic suppression of flows due to Lorentz forces. In particular, 
we find that the magnetic fields largely suppress the amplitudes of vortical motions when magnetic flux densities 
exceed about 300 G (HMI). This magnetic suppression of vorticities or helicities is identified as that arising 
from the $\alpha$-quenching action of magnetic field. This identification is found reasonable as much of the 
magnetic flux in the solar photosphere is known to be of kG strengths, which are well above the equipartition values. 

The above results have wider implications and relations to several other related phenomena pertaining to 
magneto-convective processes in the solar photosphere. The results (1) and (2) above related to the hemispheric 
pattern of fluid kinetic helicity and its relationship to magnetic field are of importance to understanding 
transfer of helicities between the fluid and magnetic field, especially in situations where reliable measurements 
of magnetic helicities in the small-scale are hard to come by. Since we have not measured magnetic helicities, 
we cannot establish the connections or transfer of helicities between fluid motions and the magnetic field. 
However, if transfer of helicities happen between fluid motions and magnetic field, irrespective of which direction 
it happens, we expect both to have the same sign. In our analysis, we find kinetic helicity is negative (positive) 
in northern (southern) hemisphere and this is consistent with magnetic helicity followed by active regions. On the 
other hand, if the small-scale fields have opposite sign of current helicity as compared to active regions as 
reported by \citet{Sanjay2013}, then it could not have gone from fluid motions as this is not consistent with 
our observation. Thus, if the magnetic helicity of small-scale magnetic field all over the Sun has contributions 
from kinetic helicity, then we expect it to have the same sign as active regions. Hence, our results 
indicate that small-scale magnetic helicity would indeed be of same sign as that of active regions that 
follow the usual hemispheric trend. 

Transfer and redistribution of vorticities by magnetic fields implied by results (2) and (3) exemplify the 
basic magnetohydrodynamic effect known from early laboratory experiments involving conducting liquid metals 
performed and expounded by \citet{Shercliff1971}. These same laboratory experiments also demonstrate suppression 
of swirly motions by magnetic field, when the field and fluid velocity directions are not aligned. 
Our result (4) here is an example for this phenomenon on the Sun. 

As regards the relevance of our results for understanding the nature of fluid turbulence and associated 
dynamo-sustaining state of the near-surface convection \citep{Brandenburg1998, Rudiger2001}, we note that 
the dominance of negative kinetic helicity in the north, and hence the expected same sign for magnetic helicity, 
indicate a positive $\alpha$-effect in the north in contrast to that required for explaining the observed 
butterfly diagram of large-scale active regions. As referred to earlier, more detailed analyses of distribution 
of kinetic helicity, its interactions with magnetic field, its spatio-temporal spectra and their relations to 
various other local and global properties of solar magnetic fields are necessary to make progress towards 
devising observational diagnostics of possible dynamo actions happening in the near-surface layers. We note that 
continuous wide field-of-view observations of higher spatial resolution to resolve sub-granular 
scale flows and magnetic fields than provided by SDO/HMI is important to address such details. We believe the kind 
of analyses undertaken here, {\it e.g.} such as those derived from the results in Figure \ref{2dplotV}, as applied 
to magnetic regions in different dynamical state such as emerging flux regions, decaying active regions, 
and those before, during and after major atmospheric activity such as eruptions or flares, would aid in 
understanding the interactions between fluid and magnetic helicities, their evolution, exchange and transport 
upwards in the atmosphere. \\

Data-intensive numerical computations required in this work were carried out using the High Performance Computing 
facility of the Indian Institute of Astrophysics, Bangalore. We thank Ravindra, B. for help with running the FLCT code. 
This work has utilised extensively the HMI/SDO data pipeline at the Joint Science Operations Center (JSOC), Stanford 
University. We thank the JSOC team at Stanford Solar Observatories Group. Our thanks are also due to an anonymous referee
for a number of constructive comments and suggestions that improved the discussion of results and the presentation in this paper.


\begin{thebibliography}{}

\bibitem[Balmaceda et al.(2010)]{Balmaceda2010}
Balmaceda L., Dominguez V. S., Palacios J., et al. 2010, A\&A, 513, 
L6. 1004.1185

\bibitem[Berger \& Field(1984)]{Berger1984}
Berger M. A., \& Field G. B. 1984, J. Fluid Mech., 147,133 

\bibitem[Bonet et al.(2008)]{Bonet2008}
Bonet J. A., Marquez I., Sanchez Almeida J., et al. 2008, ApJL, 687, 131

\bibitem[Brandenburg et al.(1995)]{Brandenburg1995}
Brandenburg A., Nordlund A., Stein R. F., \& Torkelsson U. 1995, \apj, 446, 741

\bibitem[Brandenburg \& Schmitt(1998)]{Brandenburg1998}
Brandenburg A., \& Schmitt D. 1998, A\&A, 338, L55

\bibitem[Brandenburg \& Subramanian(2005)]{Brandenburg2005}
Brandenburg A., \& Subramanian K. 2005, Physics Reports, 417, 1

\bibitem[Brandt et al.(1988)]{Brandt1988}
Brandt P. N., Scharmer G. B., Ferguson S., et al. 1988, Nature, 335, 238

\bibitem[Duvall \& Gizon(2000)]{Duvall2000}
Duvall Jr. T. L., \& Gizon L. 2000, Solar Physics, 192, 177

\bibitem[Gizon \& Duvall(2003)]{Gizon2003}
Gizon L., \& Duvall Jr. T. L. 2003, ASP, 517, 43

\bibitem[Gizon \& Birch(2005)]{Gizon2005}
Gizon L., \& Birch A.C. 2005, Living Rev. Solar Phys., 2, 6

\bibitem[Goode et al.(2010)]{Goode2010}
Goode P.R., Yurchyshyn V., Cao W., et al. 2010, \apj, 714, L31

\bibitem[Gosain S. et al.(2013)]{Sanjay2013}
Gosain S., A. A. Pevtsov, G. V. Rudenko, S. A. Anfinogentov 2013, \apj, 772, 52

\bibitem[Hale G. E.(1927)]{Hale1927}
Hale G. E. 1927, Nature, 119, 708

\bibitem[Hindman et al.(2009)]{Hindman2009}
Hindman B. W.,Haber D. A., \& Toomre J., 2009, \apj, 698, 1749

\bibitem[Hoeksema et al.(2014)]{Hoeksema2014}
Hoeksema J. D., Liu Y., Hayashi K., et al. 2014, Solar Physics, 289, 3483

\bibitem[Innes et al.(2009)]{Innes2009}
Innes D. E., A. Genetelli, R. Attie, \& H. E. Potts 2009, A\&A, 495, 319

\bibitem[Komm et al.(2007)]{Komm2007}
Komm R., Howe R., Hill F., et al. 2007, \apj, 667, 571

\bibitem[Komm et al.(2014)]{Komm2014}
Komm R., Gosain S., Petsov A. 2014, Solar Physics, 289, 475

\bibitem[Krause and R{\"a}dler(1980)]{KrauseRadler80} Krause, F. \& R{\"a}dler, K.-H. 1980, 
{\it Mean-field Magnetohydrodynamics and Dynamo Theory}, Akademie-Verlag, Berlin, and Pergamon Press, Oxford

\bibitem[Langfellner et al.(2015)]{Langfellner2015}
Langfellner J., Gizon L., \& Birch A. C. 2015, A\&A, 581, 67

\bibitem[Lisle \& Toomre(2004)]{Lisle2004}
Lisle J., \& Toomre J. 2004, Proceedings of SOHO 14/GONG 2004 workshop, ESA SP 559, 556

\bibitem[Lites et al.(2008)]{Lites2008}
Lites B.W., Kubo M., Socas-Navarro H., et al. 2008, \apj, 672, 1237

\bibitem[Liu et al.(2014a)]{Liu2014a}
Liu Y., Hoeksema J. T., \& Sun X. 2014,\apjl, 783, L1

\bibitem[Liu et al.(2014b)]{Liu2014b}
Liu Y., Hoeksema J. T., Bobra M., Hayashi K., Schuck P. W., \& Sun X. 2014,\apj, 785, 13

\bibitem[Longcope et al.(1998)]{Longcope1998}
Longcope G. H., Fisher G. H., \& Pevtsov A. A. 1998, Geophysical Monographs Series, 111, 93

\bibitem[Moffat (1978)]{Moffat78}
Moffat, H.K. 1978, {\it Magnetic Field Generation in Electrically Conducting Fluids}, Cambridge University Press, Cambridge.

\bibitem[Nordlund et al.(2009)]{Nordlund2009}
Nordlund A., Stein R. F., \& Asplund, M. 2009, Living Rev. Solar Phys., 6, 2; http://www.livingreviews.org/lrsp-2009-2

\bibitem[November \& Simon(1988)]{November1988}
November L. J., \& Simon G. W. 1988, \apj, 333, 427

\bibitem[Parker (1955)]{Parker55}
Parker, E.N. 1955, \apj, 121, 491

\bibitem[Pevtsov et al.(1995)]{Pevtsov1995}
Pevtsov A. A., Canfield R. C, \& Metcalf T. R. 1995, ApJL, 440, 109

\bibitem[Proctor \& Weiss(1982)]{Proctor1982}
Proctor M. R. E., \& Weiss N. O. 1982, Reports on Progress in Physics, 45, 11

\bibitem[Rajaguru et al.(2013)]{Rajaguruetal2013}
Rajaguru, S. P., Couvidat, S., Sun, Xudong, Hayashi, K., \& Schunker, H. 2013, Solar Physics, 287, 107

\bibitem[Rogachevskii \& Kleeorin(2000)]{Rogachevskii2000}
Rogachevskii I., \& Kleeorin N. 2000, Physical Review E, 61, 5202

\bibitem[Rudiger et al.(1999)]{Rudiger1999}
Rudiger G., Brandenburg A., \& Pipin V. V. 1999, Astron. Nachr.320, 3, 135

\bibitem[Rudiger(2001)]{Rudiger2001}
Rudiger G. 2001, IAU Symposium, 203, 152

\bibitem[Rudiger et al.(2001)]{Rudigeretal2001}
Rudiger G., Pipin V. V., \& Belvedere G. 2001, Solar Physics, 198, 241

\bibitem[Scherrer et al.(2012)]{Scherrer2012}
Scherrer P. H., Schou J., Bush R. J., et al. 2012, Solar Physics, 275, 207

\bibitem[Schou et al.(2012)]{Schou2012}
Schou J., Scherrer P. H., Bush R. I. et al. 2012, Solar Physics, 275, 229

\bibitem[Schussler \& Vogler(2008)]{Schussler2008}
Schussler M., \& Vogler A. 2008, A\&A, 481, L5

\bibitem[Seehafer (1990)]{Seehafer90}
Seehafer, N. 1990, Solar Physics, 125, 219

\bibitem[Shelyag et al.(2013)]{Shelyag2013}
Shelyag S., Cally P. S., Reid A., \& Mathioudakis 2013, \apjl, 776, 4

\bibitem[Shercliff(1971)]{Shercliff1971} 
Shercliff, J.A. 1971, {\it Film Notes for Magnetohydrodynamics, National
Committee for Fluid Mechanics Films}, Education Development Center, Inc., MIT, USA.

\bibitem[Simon et al.(1989)]{Simon1989}
Simon G. W., November L. J., Ferguson S. H., et al. 1989, NATO Advanced Science Institutes (ASI) Series C, 371

\bibitem[Wang et al.(1995)]{Wang1995}
Wang Y., Noyes R. W., Tarbell T. D., \& Title A. M. 1995, \apj, 447, 419

\bibitem[Wedemeyer-Bohm et al.(2012)]{Bohm2012}
Wedemeyer-Bohm S., Scullion E., Steiner O., et al. 2012, Nature, 486, 505

\bibitem[Welsch et al.(2004)]{Welsch2004}
Welsch B. T., Fisher G. H., \& Abbett W.P. 2004, \apj, 610, 1148

\bibitem[Welsch et al.(2007)]{Welsch2007}
Welsch B. T., Abbett W. P., De Rosa M. L., et al. 2007, \apj, 670, 1434

\bibitem[Zhang et al.(2006)]{ZhangH2006}
Zhang H., Sokoloff D., Rogachevskii I., et al. 2006, MNRAS, 365, 276

\bibitem[Zhang(2006)]{Zhang2006}
Zhang M. 2006, ApJL, 646, 85

\end{thebibliography}
\end{document}